# ALL-OPTICAL SWITCHING USING THE QUANTUM ZENO EFFECT AND TWO-PHOTON ABSORPTION

B.C. Jacobs
The Johns Hopkins University, Applied Physics Laboratory, Laurel, MD 20723

J.D. Franson
University of Maryland, Baltimore County, Baltimore, MD 21250

*Abstract:*

We have previously shown that the quantum Zeno effect can be used to implement quantum logic gates for quantum computing applications, where the Zeno effect was produced using a strong two-photon absorbing medium. Here we show that the Zeno effect can also be used to implement classical logic gates whose inputs and outputs are high-intensity fields (coherent states). The operation of the devices can be understood using a quasi-static analysis, and their switching times are calculated using a dynamic approach. The two-photon absorption coefficient of rubidium vapor is shown to allow operation of these devices at relatively low power levels.

## I. INTRODUCTION

In the quantum Zeno effect [1], a randomly-occurring process can be inhibited by frequent measurements to determine whether or not it has occurred. We have previously shown [2-4] that the quantum Zeno effect can be used to suppress the failure events that would otherwise occur in a controlled-NOT (CNOT) quantum logic gate based on a linear optics approach [5, 6]. Other groups [7, 8] have shown that Zeno logic gates of this kind may have advantages over a purely linear optical approach, although stringent requirements on the performance of the devices must be met in order to satisfy the error correction threshold for quantum computing.

Although the quantum Zeno effect was originally based on the use of frequent measurements, it was soon realized that equivalent effects could be obtained using any form of decoherence or dissipation that prevents the growth of specific probability amplitudes. For example, the Zeno quantum logic gates that we proposed earlier are based on the use of strong two-photon absorption to suppress the probability amplitude that two photons will emerge from the same output port, which is the only intrinsic source of failures in the CNOT gate of Ref. [6].

Here we show that similar effects can also be used to implement classical logic gates whose inputs and outputs correspond to intense classical beams of light (coherent states). Although many photons are involved in the operation of these devices, strong two-photon absorption can still inhibit a beam of light at a specific frequency from coupling into a resonant optical cavity. This allows the presence or absence of one beam of light to control the output path taken by a second beam of light, which corresponds to the operation of an all-optical switch.

Aside from any practical applications of these devices, these results raise some questions as to whether or not the Zeno effect is inherently quantum-mechanical in nature. It will be argued here that the Zeno effect based on dissipation or decoherence (rather than actual measurements) is a possibility in any system described by a wave equation and it is not limited only to quantum mechanics.

The paper begins with a basic description of classical logic gates based on the Zeno effect in Section II, which also includes a quasi-static analysis of their operation. These calculations assume that a toroidal microresonator [9] is used to enhance the intensity of the field inside the resonator as compared to that in the coupling waveguides. Section III discusses the all-optical switching and memory operations that can be implemented in this way, including a dynamic calculation that gives their associated switching times. These devices require a strong two-photon absorbing medium to produce the required dissipation, and Section IV calculates the two-photon absorbing properties of rubidium vapor using a quantum-mechanical approach. Further details of the performance analysis are discussed in Section V, while a summary and conclusions are discussed in Section VI.

## II. QUASI-STATIC ANALYSIS

At moderate intensities, the operation of logic gates based on the quantum Zeno effect can be understood using a classical analysis. A quantum-mechanical calculation using discrete numbers of photons would be required at sufficiently low intensities, but quantum effects become negligible and Maxwell's equations can be used when the mean number of photons is much larger than one,. This approach is still valid at intensities of picowatts or less. Although the fields are treated classically here, a quantum-mechanical description of the two-photon absorption process in an atomic vapor is given in Section IV.

We will first consider a quasi-static situation in which the input fields are slowly-varying compared to the relevant time constants associated with the resonant cavities and the two-photon absorption process. A dynamic calculation that can be used to evaluate the switching times associated with pulsed inputs will be discussed in Section III. The quasi-static analysis is relatively straightforward and it gives estimates of the losses and other potential error sources under slowly-varying conditions. It also illustrates the basic mode of operation of the devices and the bistable nature of their output.

The quasi-static treatment of the logic devices is somewhat analogous to the usual classical description of the response of a resonant cavity. The system of interest is shown in Fig. 1. A toroidal resonator is coupled to two input fields using two tapered fibers or waveguides. The electric field in either input is coupled into the toroid with a coupling coefficient $iR$, while the transmission coefficient $T$ describes the transmission of the field through the coupling regions. (This notation is meant to be analogous to the reflection and transmission coefficients of a beam splitter.) The factor of $i$ in the coupling coefficient allows $R$ and $T$ to be taken to be real numbers.

Classical electric fields $E_{1A}$ and $E_{1B}$ with angular frequencies $\omega_1$ and $\omega_2$ are assumed to be incident in paths A and B. To simplify the analysis, the coupling coefficients $R$ and $T$ are assumed to be the same in both paths, although that need not be the case in general. The coupling regions are assumed to be sufficiently short that we can ignore any losses in those regions. The electric fields will be assumed to decay at a rate $\gamma$ per cm of travel due to linear (single-photon) losses that may include scattering by the atoms as well as the intrinsic loss mechanisms of the cavity. In addition, each electric field $E_i$ inside the resonator is assumed to decay at a rate of $\alpha I_j$ per cm of travel, where $\alpha$ is the usual two-photon absorption coefficient and $I_j = E_j^* E_j$ is the intensity of the other field. The quality factor $Q$ for the resonator will be assumed to be sufficiently high that the intensities $I_j$ in this expression are essentially constant throughout the resonator.

It will be assumed that two-photon absorption can only occur for two photons of different frequency and that two photons of the same frequency cannot be absorbed. It will be shown in Section IV that the self two-photon absorption can be up to eight orders of magnitude smaller than the cross two-photon absorption when rubidium vapor is used as the atomic medium.

Under quasi-static conditions, the field at frequency $\omega_1$ can be determined at various points around the resonator as illustrated in Fig. 2. For now, we will assume that the intensity of the other field has a known value of $I_2$ for the purpose of computing the losses due to two-photon absorption. We will define the value of the field at frequency $\omega_1$ just after the input coupling as $E_{1R}$, as indicated by one of the arrows in the figure. After the field has propagated half-way around the toroid to just before the other coupling region, its amplitude $E_{1R}'$ will be given by

$$E_{1R}' = E_{1R} e^{ik_1 L/2} e^{-\gamma L/2} e^{-\alpha I_2 L/2} \tag{1}$$

where L is the circumference of the toroid. Eq. (1) includes the losses due to single-photon and two-photon absorption, as well as a phase shift corresponding to the effective propagation constant $k_1$ of the field inside the toroid.

After passing through the second coupling region, the electric field inside the toroid will be reduced by the transmission coefficient to

$$E_{1R}'' = E_{1R} e^{ik_1 L/2} e^{-\gamma L/2} e^{-\alpha I_2 L/2} T \tag{2}$$

Further propagation to the point just before the upper coupling region produces a field given at that point by

$$E_{1R}''' = E_{1R} e^{ik_1 L} e^{-\gamma L} e^{-\alpha I_2 L} T \tag{3}$$

Finally, the field $E_{1R}$ just after the upper coupling region must be a superposition of the incident field $E_{1A}$ multiplied by the coupling coefficient $iR$ plus the field $E_{1R}'''$ multiplied by the transmission coefficient $T$. This gives the consistency condition

$$E_{1R} = E_{1R} e^{ik_1 L} e^{-\gamma L} e^{-\alpha I_2 L} T^2 + iR E_{1A} \tag{4}$$

which must be satisfied under quasi-static conditions.

Solving Eq. (4) for the value of $E_{1R}$ gives the result that

$$E_{1R} = i \frac{R}{1 - e^{ik_1 L} e^{-\gamma L} e^{-aI_{2R}L} T^2} E_{1A} \tag{5}$$

Here we have made the approximation that $I_2 \doteq I_{2R} = E_{2R}^* E_{2R}$ as discussed previously for a high-Q cavity. In a similar way, it can be shown that the electric field at frequency $\omega_2$ inside the resonator and just after the lower input coupling is given by

$$E_{2R} = i \frac{R}{1 - e^{ik_2 L} e^{-\gamma L} e^{-aI_{1R}L} T^2} E_{2A} \tag{6}$$

In this case, it was assumed that the intensity of the field at the other frequency $\omega_1$ had a known intensity $I_{1R} = E_{1R}^* E_{1R}$.

Eq. (5) gives the value of $E_{1R}$ as a function of the assumed value of $I_2 \doteq E_{2R}^* E_{2R}$, while Eq. (6) gives the value of $E_{2R}$ as a function of the assumed value of $I_1 \doteq E_{1R}^* E_{1R}$. The solutions to this set of coupled nonlinear equations determines the quasi-static response of the system. The form of the solutions can be complicated, however, since they are nonlinear.

Some insight into the nature of the solutions can be obtained by comparing the plots of these equations in Figs. 3 through 5. Figure 3 shows the calculated value of $I_{2R}$ as a function of the assumed value of $I_{1R}$, while Fig. 4 shows the calculated value of $I_{1R}$ as a function of the assumed value of $I_{2R}$. Both fields were assumed to be on resonance with the cavity and the assumed parameters of the resonator are given in Section V. These two curves are plotted on top of each other near the origin in Fig. 5, where they intersect at the point where $I_{1R} = I_{2R}$. The intersection of these two curves corresponds to a symmetric solution to Eqs. (5) and (6).

The solutions to these coupled equations can be found numerically using an iterative approach. We begin by selecting a trial value for $I_{2R}$ and inserting it into Eq. (5) to determine the corresponding value of $E_{1R}$ and $I_{1R} \doteq E_{1R}^* E_{1R}$. This value of $I_{1R}$ can then be inserted into Eq. (6) to determine a revised value for $I_{2R}$. By repeating this process multiple time, $I_{1R}$ and $I_{2R}$ will rapidly converge to a self-consistent set of values. The results of this numerical calculation were in excellent agreement with the dynamic calculations described in Section III.

The iterative calculations showed that the symmetric solution with $I_{1R} = I_{2R}$ is actually unstable. Given a starting value arbitrarily close to that point, the iterative approach will rapidly move away and converge to one of two bistable solutions in which either $I_{1R} \gg I_{2R}$ or $I_{1R} \ll I_{2R}$. These solutions correspond to a situation in which the intensity of one of the beams is so low that it produces negligible two-photon absorption at the other frequency. In that case, the other beam will be strongly coupled into the cavity and will produce a large intensity on resonance. The field with the high intensity will then produce strong two-photon absorption at the other frequency, which ensures that the low-intensity beam will continue to be only weakly coupled into the cavity.

The bistability of the solutions with $I_{1R} \gg I_{2R}$ or $I_{1R} \ll I_{2R}$ can be understood from the fact that the two-photon absorption produced by the weaker field is very small, so that small fluctuations in its intensity do not significantly affect the solution for the stronger field. These results from the iterative approach were found to be in good agreement with a formal stability analysis.

Given a consistent solution for the fields $E_{1R}$ and $E_{2R}$ inside the resonator, the output fields that propagate away from the toroid can be determined. An output field $E_{1A}'$ at frequency $\omega_1$ will be produced in waveguide A and it will propagate towards the right, as shown in Fig. 1. The output field $E_{1A}'$ is a superposition of one component that is transmitted through the coupling region and a second component coupled out of the resonator. From Fig. 2, this can be seen to give

$$E_{1A}' = TE_{1A} + iRE_{1R}e^{ik_1 L}e^{-(\gamma+\alpha I_{1R})L}T \tag{7}$$

A similar expression exists for the field $E_{1B}'$ at frequency $\omega_1$ that is coupled out of the resonator and into waveguide B, which propagates towards the left (not shown in the figure). The corresponding fields $E_{2B}'$ and $E_{2A}'$ at frequency $\omega_2$ can be calculated in a similar way. These expressions can be used to determine the output of the device at both frequencies and in both wave guides.

## III. SWITCHING AND MEMORY OPERATIONS

The bistable solutions with $I_{1R} \gg I_{2R}$ or $I_{1R} \ll I_{2R}$ allow devices of this kind to function as an all-optical switch. Suppose that the input field in waveguide A is initially turned on while the input field in waveguide 2 is turned off. The field in waveguide A will then establish a large intensity inside the resonator at frequency $\omega_1$. If the input field in waveguide B is subsequently turned on, strong two-photon absorption will prevent it from being coupled into the resonator. In that case, nearly all of the power at frequency $\omega_2$ will be transmitted straight through the device and continue to propagate in waveguide B.

Now suppose that the input field in waveguide A was initially turned off. If the input field in waveguide B is subsequently turned on, it will experience no two-photon absorption and it will be strongly coupled into the resonator. For an appropriate value of the coupling coefficient R (critical coupling), nearly all of the incident power at frequency $\omega_2$ will be transferred into waveguide A and travel to the left. Thus the initial intensity in waveguide A can be used to switch the power in waveguide B into one of two output paths, which corresponds to the operation of an all-optical switch.

The bistable nature of these solutions also allows the devices to be used as a all-optical memory device. In this case the condition in which $I_{1R} \gg I_{2R}$ could be taken to represent a bit value of 0, while the condition in which $I_{1R} \ll I_{2R}$ could represent a bit value of 1. The value of the bit can be set or reset using an appropriate sequence of quasi-static signals, and the bistability of the solutions ensures that the memory will be stable over an indefinitely large period of time.

These effects are similar to other bistable phenomena in optical materials, which can also be used to implement switching or logic operations. The devices reported here differ, for example, in that two-photon absorption provides the necessary nonlinearity.

One measure of the performance of these switching and memory devices is the fraction of the power that is transmitted into the two possible output paths for a given intensity in the input paths. Ideally, one would like to have all of the input power switched into the desired output path with no power transmitted into the other path. The actual performance will be limited by losses in the cavity as well as the magnitude of the two-photon absorption coefficient. These device characteristics are calculated in Section V as a function of the known characteristics of the resonator, such as its Q value, and the two-photon absorption coefficient. The results of those calculations show that the errors in the output signals can be less than 1% for typical resonator parameters and for the two-photon absorption coefficient of rubidium vapor, as calculated in Section IV.

The quasi-static analysis of Section I provides insight into the operation of these devices and it was used to calculate the performance factors described above. When pulsed inputs are used, however, one would also like to know the characteristic switching times of the devices. This can be determined using a dynamic calculation that assumes, once again, that the Q is sufficiently high that the two-photon absorption rate is nearly constant around the circumference of the toroid. Here the time scale was divided into intervals that correspond to one transit time around the toroid. The values of the fields at the points shown in Fig. 2 were updated at each time interval using the fields that had propagated around the toroid from that point since the previous update. The results of such a calculation are shown in Fig. 6, where the dynamic response of the system to a short control pulse is plotted.

The results in Fig. 6 assumed the use of a relatively large control signal to switch a continuous input. However, in a pulsed system the control can be of comparable (or lower) intensity than the target. This is shown in Fig. 7, where a 25 µW control is used to switch a 25 µW peak signal.

It can be seen from these dynamic calculations that the switching time of the devices is determined by the storage time of the cavity as well as the rate of single-photon and two-photon absorption. For the parameters listed in Section V, the typical switching times are on the order of 200 ps with input intensities as low as 25 pW.

### IV. TWO-PHOTON ABSORPTION IN RUBIDIUM VAPOR

The operation of logic and memory devices of this kind requires a medium that produces stronger two-photon absorption than single-photon absorption. This section describes the calculation of the rate of two-photon absorption using rubidium vapor in combination with a toroidal microresonator. The rate of single-photon absorption due to scattering by the rubidium atoms is also calculated, as is the self two-photon absorption at a single frequency. These calculations indicate that sufficiently strong two-photon absorption can be achieved using rubidium vapor.

Rubidium vapor gives very high rates of two-photon absorption as compared to solid-state materials. This is because the atoms in an atomic vapor have a very narrow bandwidth, which gives a strong response to two photons whose total energy is on resonance with the upper atomic transition. Homogeneous and inhomogeneous broadening in typical solid-state materials broaden out the linewidth, which spreads the absorption out over a larger range of wavelengths and thus reduces the two-photon absorption at any given wavelength. For similar reasons, rubidium vapor gives much lower single-photon and self two-photon losses than do solid-state materials.

Other atomic species such as sodium could also be used to produce two-photon absorption. Most of these atoms, including rubidium, have strong dipole moments compared with solid-state materials, which also enhances the rate of two-photon absorption. Rubidium has the advantage of having two successive atomic transitions whose wavelengths differ by only 4 nm. This allows relatively small detunings, which enhances the strength of the two photon absorption. In addition, the two wavelengths are both near 780 nm, which is a convenient wavelength to use in microresonators. Rubidium has the further advantage of having a large vapor pressure at temperatures near room temperature.

The rubidium vapor must be generated inside a hermetically sealed cell at some temperature T, which controls the vapor pressure and thus the density of rubidium atoms. This could be easily accomplished by sealing the resonators inside glass envelopes similar to vacuum tubes. Several trade-offs exist with regard to the optimal operating temperature and atomic density, as will be described below.

*A. Perturbation theory calculation of the two-photon absorption rate*

Because of the narrow linewidths associated with rubidium vapor, the rate of two-photon absorption is a strong function of the wavelengths of the two beams of light. This is especially true if one of the wavelengths is detuned by a relatively small amount from the resonant frequency of the first atomic transition. As a result, there is no generic value for the two-photon absorption coefficient $\alpha$ for rubidium vapor and the coefficient cannot be determined from tabulated values.

Instead, the rate $R_2$ of two-photon absorption was calculated from basic principles using quantum-mechanical perturbation theory and density operator techniques [10]. The atomic system of interest is illustrated in the "ladder" diagram of Fig. 8. Rubidium vapor has three atomic levels labeled 1, 2, and 3, which correspond to the $5\,{}^2S_{1/2}$, $5\,{}^2P_{3/2}$, and $5\,{}^2D_{5/2}$ states, respectively. As a result, the dipole selection rules allow strong transitions between these three levels. However, the frequencies of the two beams do not match the resonant frequency of the transition from levels 1 and 2, as illustrated in the figure. As a result, a single photon cannot be absorbed.

But the sum of the two frequencies is chosen to match the total change in the energy of the atom between levels 1 and 3, so that the absorption of two photons does conserve energy. The rate of two photon absorption depends in part on the difference $\delta$ between the energy of photon 1 and the first atomic transition. Energy need not be conserved over small time intervals in quantum mechanics, and the first atomic transition is referred to as a virtual quantum state.

Calculations involving the absorption of two photons in this way can be performed using density operator techniques [10]. If the rate of absorption is sufficiently small that the initial state is not significantly depleted, as is expected to be the case in the operation of these devices, then density operator calculations give

$$R_2 = \frac{8\left|\langle 3|q\mathbf{r}\cdot\mathbf{E}|2\rangle\langle 2|q\mathbf{r}\cdot\mathbf{E}|1\rangle\right|^2}{\hbar^2 \delta^2} \frac{1}{\gamma_2} \tag{8}$$

for the rate of two-photon absorption by a single atom. Here q is the charge of an electron, $\mathbf{r}$ is the location of the electron within an atom, $\mathbf{E}$ is the electric field operator at the location of the atom, $\hbar$ is Planck's constant divided by $2\pi$, and $\gamma_2$ is the half-width of atomic level 3 due to collisions and spontaneous emission. Eq. (8) is consistent with the formula for single-photon absorption given in Ref. (11) with the effective matrix element for two-photon absorption substituted for that associated with single-photon absorption. It is also consistent with the results of Ref. (4).

The matrix elements $\langle j|\mathbf{r}|i\rangle$ are randomly distributed in direction and are equal in magnitude to the dipole moment of the associated atomic transition. The dipole moments can be calculated from the tabulated lifetimes of the corresponding atomic states, and their values are $2.23\times 10^{-10} m$ and $0.492\times 10^{-10} m$ for the first and second transitions, respectively.

The magnitude of the electric field will be a function of the position of each atom with respect to the center of the toroidal resonator. For the purpose of these calculations, the resonator was assumed to have a major diameter of $D = 50\mu m$ and a minor diameter of $d = 0.35\mu m$. To a first approximation, the electric field distribution for a toroidal resonator with $d \ll D$ is the same as that of a straight section of optical fiber with the corresponding diameter and an index of refraction of $n = 1$ outside of the silica core. An exact formula for the field of such a silica fiber has been given by Mazur's group [12]. Vahala's group [13] has given a perturbative correction to the straight-fiber formulas for the case when the ration of $d$ to D cannot be neglected. In our case, $d/D = 0.007$ and we used only the first-order correction to the straight-fiber fields.

The two photon absorption rate was then calculated for a total field energy corresponding to that of a single photon at each frequency. The results of Eq. (8) were numerically integrated over all possible locations of an atom outside of the microresonator using Mathematica. The baseline case assumed a uniform density $\rho_0$ of $10^{14}$ rubidium atoms per cubic centimeter, a wavelength detuning $\lambda_\delta$ of 2.12 nm, and a linewidth of $\gamma_2 = 3.14\times 10^8 /\sec$. This gave a two-photon absorption rate of $R_{20} = 9.41\times 10^8 /\sec$, where the subscript 0 refers to the "baseline" parameters. A single-photon loss rate of $R_{10} = 1.12\times 10^8 /\sec$ was obtained under the same conditions. Thus the two-photon absorption rate is larger than the single-photon absorption rate even at single-photon intensities; the ratio is even higher at classical intensities, as described below.

For conditions different from the baseline parameters described above, the two-photon and single-photon absorption rates scale approximately as

$$R_2 = \frac{\rho}{\rho_0}\left(\frac{\delta_0}{\delta}\right)^2 \frac{I_1 I_2}{I_{10} I_{20}} R_{20}$$
$$R_1 = \frac{\rho}{\rho_0}\left(\frac{\delta_0}{\delta}\right)^2 \frac{I_1}{I_{10}} R_{10}.$$
(9)

Here $\delta_0$ is the baseline detuning of 2.12 nm, $I_1$ and $I_2$ are the intensities of the two beams inside the resonator, and $I_{10}$ and $I_{20}$ are the intensities corresponding to a single photon in the resonator. These relations will be used to perform a variety of trade-off studies below.

*B. Effective two-photon absorption coefficient*

The performance estimates for solid-state materials were obtained using a two-photon absorption coefficient $\alpha$ with units of cm/GW. For comparison purposes, the two-photon absorption rate determined above can be converted to an effective value of $k_2$ by calculating the effective cross-sectional area A of the toroidal resonator mode structure. The calculated mode volume V of the toroidal resonator was $7.6\times 10^{-11} cm^3$ and its circumference is $\pi D$. This gives an effective area of

$$A = V/\pi D = 4.83\times 10^{-9} cm^2.$$
(10)

The average energy flux $P_f$ (in joules per second) corresponding to a single photon circulating in the resonator is given by its energy divided by the round trip travel time $\Delta t$ around the toroid:

$$P_f = \frac{\hbar\omega}{\Delta t}.$$
(11)

The round-trip travel time is given by

$$\Delta t = \frac{\pi D}{c/n_{eff}}.$$
(12)

Here c is the free-space speed of light and $n_{eff}=1.30$ is the effective index of refraction, which corresponds approximately to a weighted average of the index of refraction of silica and $n=1$ outside of the core; the weighting factors were based on the fraction of the energy in the evanescent field outside the core. This gives $\Delta t = 6.81\times 10^{-13}$ sec. The energy of a single photon at 778 nm is

$$\hbar\omega = \frac{hc}{\lambda} = 2.55\times 10^{-19} J.$$
(13)

Combining these three equations gives $P_f = 3.74 \times 10^{-7}$ Watts $= 3.74 \times 10^{-16}$ GW for a single photon in the resonator.

The baseline two-photon absorption rate $R_{20}$ is related to the baseline loss $L_0$ per cm of travel by

$$L_0 = \frac{R_{20}}{c/n_{eff}} = 4.08 \times 10^{-2} / cm. \tag{14}$$

The effective value of the two-photon absorption coefficient for the nominal conditions is then given by

$$\alpha_0 = \frac{L_0 A}{P_f} = 5.27 \times 10^5 \, cm/GW \tag{15}$$

The two-photon absorption coefficient for different values of $\delta$ and $\rho$ scales the same way as in Eq. (2).

*C. Other losses*

The previous two sections described the calculation of the rate of two-photon absorption, which is desired in order to enable the operation of logic and memory devices. There will also be undesirable losses, however, due to a number of mechanisms.

The linear losses due to single-photon absorption and scattering were already mentioned above. A perturbation theory calculation similar to that for the two-photon absorption gives the following formula for the rate of single-photon absorption by a single atom

$$R_1 = \frac{2|\langle 2|q\mathbf{r}\cdot\mathbf{E}|1\rangle|^2}{\delta^2 + (\hbar\gamma_1)^2} \gamma_1 \tag{16}$$

As before, the absorption rate for a single atom was integrated over the volume outside of the core of the resonator using the expressions for the electric field and a nominal density of $\rho_0$. This gave a baseline single-photon loss rate of $R_{10} = 1.12 \times 10^8$.

There will also be an intrinsic linear loss inside the resonator due to its finite value of Q. Most of this loss will probably be due to scattering from the imperfect surface of the silica. In any event, the optimal design will probably consist of adjusting the density of rubidium atoms to match the intrinsic resonator loss to the single-photon loss due to the atoms. For the purpose of these calculations, it was assumed that the total Q of the cavity due to both sources of loss was $Q = 5 \times 10^7$, which is a realistic figure for resonators of this kind.

Finally, there will be a very small amount of nonlinear loss due to the absorption of two photons with the same frequency. Although this does not appear to conserve energy, the upper atomic level 3 will have some spectral width due to collisions or other damping effects. In essence, the collisions make the energy of the upper level slightly uncertain so that there is some probability to absorb two photons whose total energy is off resonance. Including the effects of the spectral width, Eq. (8) for the rate $R_S$ of absorption of two photons with the same frequency $\omega_1$ becomes

$$R_S = \frac{8|\langle 3|q\mathbf{r}\cdot\mathbf{E}|2\rangle\langle 2|q\mathbf{r}\cdot\mathbf{E}|1\rangle|^2}{\delta^2} \frac{\gamma_2}{[2\hbar\omega_1 - (E_3 - E_1)]^2 + (\hbar\gamma_2)^2} \tag{17}$$

A similar expression exists for the absorption of two photons of frequency $\omega_2$. Figure 9 shows the ratio of the self two-photon absorption rate $R_S$ to the cross two-photon absorption rate $R_2$ as a function of the difference in the two wavelengths in nm. It can be seen that the self two-photon absorption is a factor of $10^{-8}$ smaller than the intended cross two-photon absorption for wavelength differences of 0.5 nm or more. Since that condition is expected to be satisfied in actual applications, the absorption of two photons at the same frequency can be completely neglected when using rubidium vapor. That may not be the case, however, using solid-state materials, since their linewidths are much larger.

*D. Required density and temperature*

The "baseline" calculations described above were carried out for an assumed density of $\rho_0 = 10^{14}$ atoms per cubic centimeter and a detuning of 2.12 nm. In rubidium vapor, that detuning corresponds to $\omega_1 = \omega_2$, which is convenient in some applications such as quantum computing. Here we would like to operate at a smaller detuning in order to reduce the required density and therefore the required temperature of the rubidium cell. A larger difference between the two wavelengths also reduces the self two-photon absorption, as discussed above.

The vapor pressure P of rubidium (in torr) depends on the temperature T (in degrees Kelvin) as given by the approximate formula

$$\log_{10} P = 15.88253 - 4529.635/T + 0.00058663 \times T - 2.99138 \times \log_{10}(T) \tag{18}$$

This can be converted to the density $\rho$ of atoms per cubic centimeter by using the ideal gas law:

$$\rho = P \times 9.63 \times 10^{18}/T. \tag{19}$$

These two equations can be combined to calculate the temperature required to produce a given density.

From the scaling dependence given in Eq. (2), it can be seen that the nominal rate $R_{20}$ of two-photon absorption can be obtained at a lower density if the detuning is decreased from the nominal value of 2.12 nm. Figure 10 shows a plot of the required rubidium density (per cm$^3$) as a function of the detuning in nm, as calculated using Eq. (2). It can be seen that densities of less than $10^{11}$ per cc will be required if the detuning is less than 0.1 nm.

Eqs. (11) and (12) can be used to calculate the temperature required to produce the rubidium density of Fig. 10. Fig. 11 plots the required temperature (in degrees C) as a function of the detuning in nm. Although there are many trade-offs to be considered, it can be seen from Fig. 11 that a detuning of 0.05 nm would allow operation at a temperature of 43 degrees C, and this appears to be the optimal temperature for the operation of the device.

## V. CALCULATION OF SWITCHING PERFORMANCE

As discussed in the main text, the potential performance of these switching and memory devices can be characterized by the amount of power that is transmitted into the wrong output port, as well as the reduction in power that is transmitted into the desired output port. In this appendix, the performance of the devices will be determined after deriving some of the basic relations characteristic of a toroidal resonator.

The magnitude of the electric field will be larger in the resonator than in the incident waveguide by a factor of $f$, which increases the rate of two-photon absorption. The factor $f$ can be determined from Eq. (7), which relates the input and output powers at frequency $\omega_1$. At critical coupling and assuming a negligible intensity at $\omega_2$, approximately all of the power will be transferred to the other waveguide, so that $E_{1A}' = 0$. On resonance, the factor of $\exp(ik_1 L)$ will be unity and, for a high-Q cavity, we can take $T \doteq 1$ and neglect the losses to a first approximation. Eq. (7) then gives

$$E_{1A} + iRE_{1R} = 0 \tag{20}$$

Solving for the factor f gives

$$f = \frac{|E_{1R}|^2}{|E_{1A}|^2} = \frac{1}{R^2} \tag{21}$$

The value of $R$ required for critical coupling can be related to the loss in the cavity by inserting the solution for $E_{1R}$ from Eq. (6) into Eq. (7) for $E_{1A}'$:

$$E_{1A}' = 0 = E_{1A} + iR \left[ i \frac{R}{1 - e^{-\gamma L}} \right] e^{-\gamma L} E_{1A}. \tag{22}$$

Here the two-photon absorption has once again been assumed to be negligible for a small intensity at $\omega_2$ and we have used $T \doteq 1$. Solving this equation for $R$ gives the approximate result that

$$R = \sqrt{\gamma L} \tag{23}$$

for critical coupling. Inserting this value of $R$ into Eq. (21) gives

$$f = \frac{1}{\gamma L} \tag{24}$$

The value of $\gamma$ can be related to the Q of the cavity by using the definition

$$Q = 2\pi \frac{E_s}{E_d} \tag{25}$$

where $E_s$ is the energy stored in the cavity and $E_d$ is the energy dissipated per optical cycle. The loss dissipated per cycle is equal to the loss experienced in traveling a distance of one wavelength inside the resonator, so that

$$E_d = E_s \gamma \frac{\lambda}{n_e} \tag{26}$$

Here $\lambda$ is the wavelength in free space and $n_e$ is the effective index of refraction in the toroid. Combining Eqs. (25) and (26) gives

$$Q = \frac{2\pi n_e}{\lambda \gamma} \tag{27}$$

This can be solved for the linear loss factor $\gamma$:

$$\gamma = \frac{2\pi n_e}{\lambda Q} \tag{28}$$

Inserting this into Eq. (24) gives

$$f \doteq \frac{\lambda Q}{2\pi n_e L} \tag{29}$$

Eq. (29) can be used to determine the intensity enhancement factor from the usual parameters describing the toroid.

An important parameter describing systems of this kind is the incident power $P_c$ in the waveguide at which the single-photon and two-photon loss rates are equal, assuming equal intensities at each frequency. This condition is determined by the requirement that

$$\alpha \frac{(fP_c)^2}{A} = \gamma f P_c \tag{30}$$

Solving this equation gives

$$P_c = \frac{\gamma A}{\alpha f} \tag{31}$$

Some typical parameters for a toroidal resonator using rubidium vapor as a two-photon absorption medium are given in Table I. The value of the two-photon absorption coefficient was obtained from a quantum-mechanical calculation as described in Section IV. With this choice of parameters, the single-photon and two-photon absorption rates would be equal at an intensity of $P_c = 3.7 \times 10^{-7}$ watts. The ratio of the two-photon absorption rate to the single-photon rate would therefore be $10^3$ at an intensity of $P_c = 3.7 \times 10^{-4}$ watts, which was used in the quasi-static analysis of section II.

Many other implementations of all-optical switches and optical bistability have previously been demonstrated, as reviewed in Ref. [14]. Examples include Mach-Zehnder interferometers with a nonlinear phase shift produced by a Kerr nonlinearity [14], ring resonators [15] whose frequency can be shifted [16, 17], and free carrier generation in quantum wells [18]. Most of these devices operate with control powers in excess of 10 mW [14]. The two-photon absorption approach described here has several potential advantages, including a strong nonlinearity enhanced by the small mode volume of the resonators. The lack of free-carrier generation also reduces the power dissipated into the substrate. Our preliminary estimates indicate that this approach may offer high-speed switching rates at lower power levels than many of the competing approaches.

## VI. SUMMARY AND CONCLUSIONS

It has been shown that strong two-photon absorption can be used to implement all-optical logic and memory devices at classical intensities. The presence of a strong field inside a resonator at one frequency can inhibit a second field at another frequency from coupling into the resonator in the presence of a two-photon absorbing medium. This allows a beam of light at one frequency to control the output path taken by a second beam of light, which allows the implementation of logic and memory operations.

These results suggest that the Zeno effect is actually a possibility in any system described by a wave equation. Although the quantum Zeno effect was originally suggested as being based on frequent measurements, it was subsequently shown that equivalent results could be obtained using any form of dissipation or decoherence to suppress the growth of certain probability amplitudes. Since dissipative effects, such as strong two-photon absorption, can occur for classical beams of light, it is possible to generalize our earlier work on quantum logic gates based on the Zeno effect to classical logic gates operating at high intensities.

As a practical matter, the implementation of classical logic gates should be much easier than the implementation of quantum logic gates for use in a quantum computer. There is no need to keep the error rate in the devices sufficiently small to meet the threshold for quantum error correction, which greatly reduces the requirement on the operating parameters. In addition, the rate of two-photon absorption at one frequency will be proportional to the intensity at a second frequency; since the intensities are much higher in a classical logic gate, the requirements on the operating parameters are further reduced.

As described in Sections II and V, toroidal microcavities combined with rubidium vapor should allow classical logic gates of this kind to operate at relatively low intensities with small error rates. These devices have the potential advantages of low-intensity operation with fanout and reset capabilities, which have not always the case for other approaches to all-optical computing.

We would like to acknowledge S. Hendrickson and H. You [10] for their previous calculations of the field modes of the toroidal resonator. This work was supported by DARPA DSO under contract HR001-D-003.

Table 1. Nominal parameters used in the performance analysis.

| Resonator quality factor Q | $5 \times 10^7$ |
|---|---|
| Major diameter of toroid D | $50\,\mu m$ |
| Minor diameter of toroid d | $0.35\,\mu m$ |
| Effective mode area A | $4.83 \times 10^{-9}\,cm^2$ |
| Effective index of refraction $n_e$ | 1.30 |
| Wavelength 1 $\lambda_1$ | 780 nm |
| Wavelength 2 $\lambda_2$ | 776 nm |
| Detuning in intermediate state | 0.05 nm |
| Power enhancement factor $f$ | 30350 |
| Density of rubidium vapor $\rho$ | $5.6 \times 10^{10}\,/cm^3$ |
| Temperature of rubidium vapor T | 43 $C$ |
| Dipole moment of the first transition | $0.23\,nm$ |
| Dipole moment of the second transition | $0.05\,nm$ |
| Decay rate of upper level due to collisions | $10^9\,/s$ |
| Equivalent single-photon loss rate $\gamma$ | $2.13 \times 10^{-3}\,/cm$ |
| Effective two-photon loss coefficient $\alpha$ | $5.27 \times 10^5\,cm/GW$ |

**Figures**

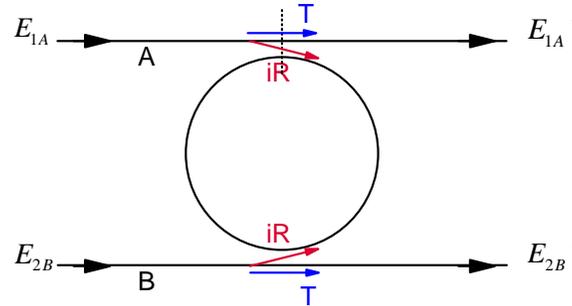

Fig. 1. Coupling of two input waveguides into a toroidal resonator with coupling coefficient $iR$ and transmission coefficient $T$. Classical electric fields $E_{1A}$ and $E_{2B}$ at frequencies $\omega_1$ and $\omega_2$ are assumed to be incident from the left. Two output fields $E_{1A}'$ and $E_{2B}'$ are shown propagating to the right. Two other output fields, $E_{1B}'$ and $E_{2A}'$, propagating to the left are not shown.

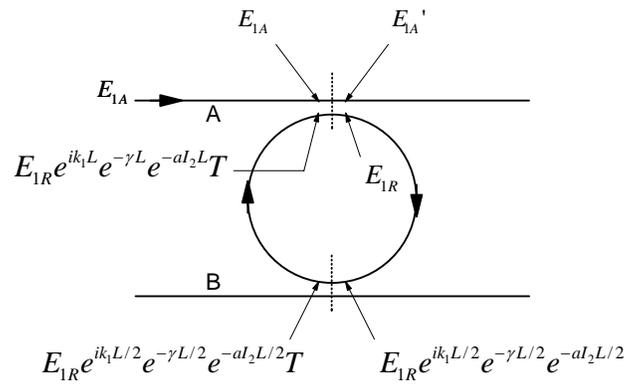

Fig. 2. Values of the electric field at frequency $\omega_1$ at various locations in the toroidal resonator, assuming a fixed value of the intensity $I_2$ of the other field for the purpose of calculating the rate of two-photon absorption.

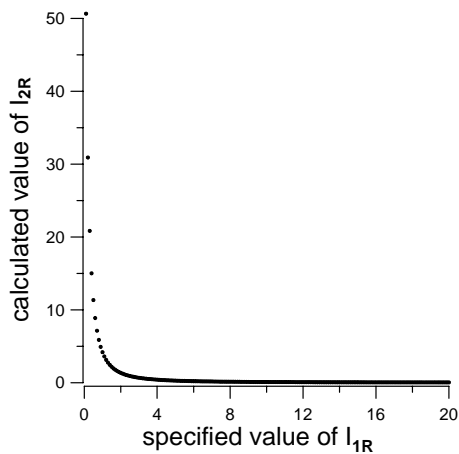

Fig. 3. Plot of the calculated value of the intensity $I_{2R}$ at frequency $\omega_2$ inside the resonator as a function of the assumed intensity $I_{1R}$ at frequency $\omega_1$.

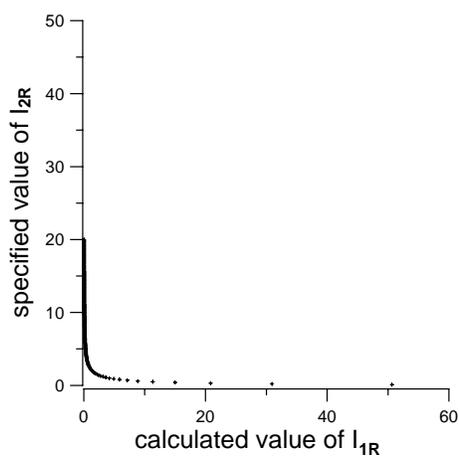

Fig. 4. Plot of the calculated value of the intensity intensity $I_{1R}$ at frequency $\omega_1$ as a function of the assumed intensity $I_{2R}$ at frequency $\omega_2$.

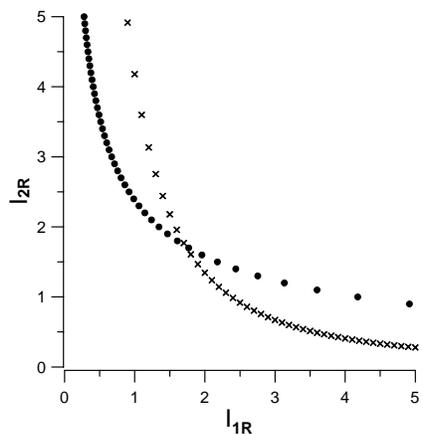

Fig. 5. The two curves from Figs. 3 and 4 superimposed near the origin. The intersection of the two curves at $I_{1R} = I_{2R}$ corresponds to one of the solutions of Eqs. (5) and (6).

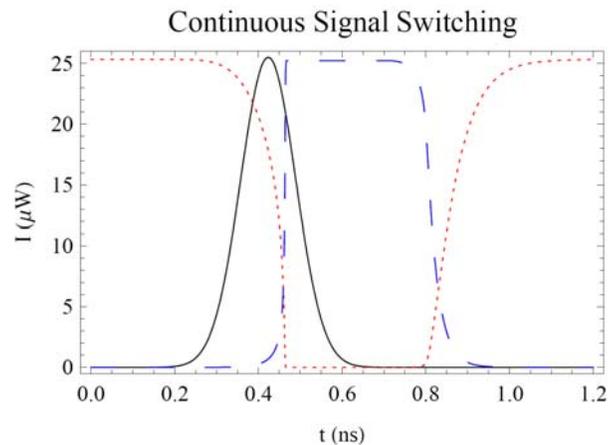

Fig. 6. Dynamic switching results, showing the ability of a control signal (black/solid and not to scale) to determine which path a constant 25 μW target signal exits the system. Initially, only the target signal is present and all its energy is in the reflected (red/dotted) mode. Within 300 ps of applying the 3mW control pulse the target field in the resonator collapses and target signal is transmitted (blue/dashed) past the resonator. Similarly, within 500 ps of turning off the control the system returns to reflecting the target input.

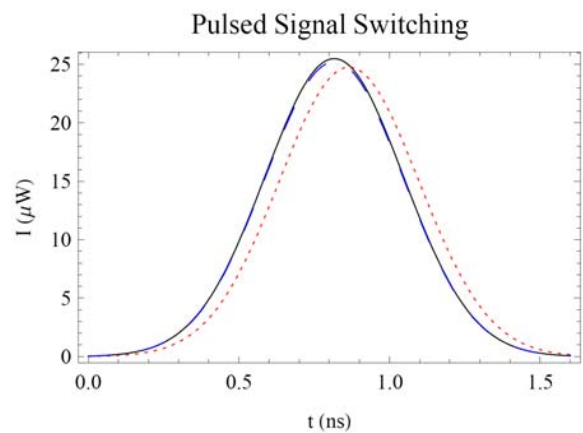

Fig. 7. Dynamic switching results for pulsed signals. Without the control applied the input target pulse (black/solid) is reflected (red/dotted) from the resonator and slightly delayed. With the 25 μW control present the target is nearly perfectly reflected (blue/dashed) with no delay.

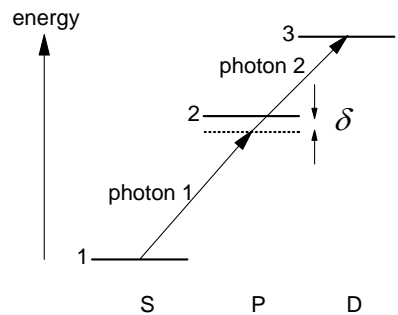

Fig. 8. Three atomic energy levels in rubidium labeled 1-3, in order of increasing energy. The letters S, P, and D refer to the angular momentum of the states in spectroscopic notation. Two photons are incident on the atom with two different frequencies that do not match that of the first atomic transition. As a result, the absorption of a single photon does not conserve energy and cannot occur, but the absorption of two photons does. The difference in energy between photon 1 and the first atomic transition (the detuning) is denoted by $\delta$.

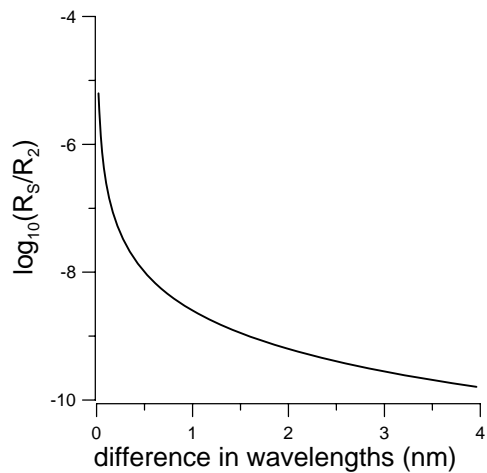

Fig. 9. Plot of the logarithm of the ratio of the self two-photon absorption rate to the cross-two photon absorption rate as a function of the difference between the wavelengths of the two beams (in nm).

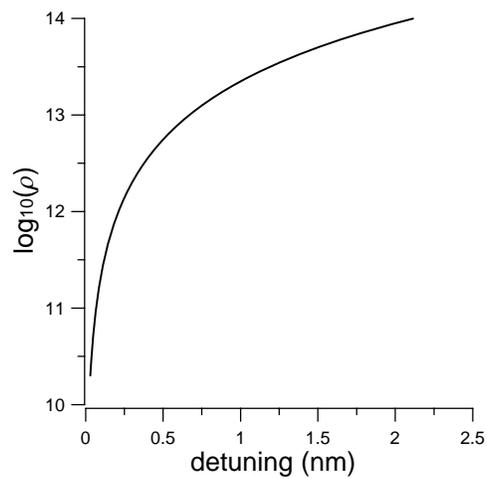

Fig. 10. Plot of the logarithm of the required rubidium density $\rho$ (in atoms/cc) as a function of the detuning in nm. This density will give the nominal two-photon absorption coefficient of $\alpha_0 = 5.27 \times 10^5$ cm/GW.

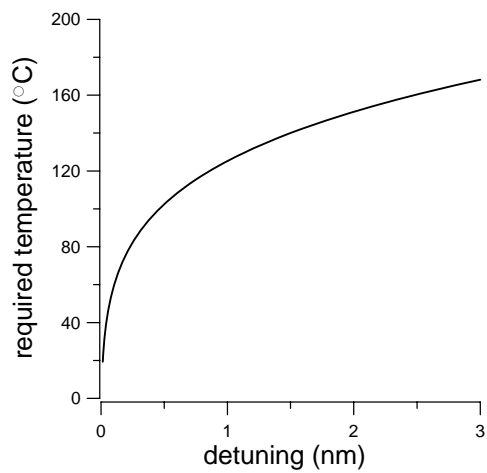

Fig. 11. Plot of the temperature (degrees C) required to achieve the nominal rate of two-photon absorption as a function of the detuning in nm.